\def\be{\begin{equation}}
\def\ee{\end{equation}}
\def\bea{\begin{eqnarray}}
\def\eea{\end{eqnarray}}
\def\bse{\begin{subequations}}
\def\ese{\end{subequations}}

\def\DT{D_{\text{T}}}
\def\perpgradT{({\hat{\bm k}}_{\perp}\!\!\cdot\!{\bm\nabla}T)}

\documentclass[sn-mathphys]{sn-jnl}
\usepackage{color}
\usepackage{bm}
\usepackage{ulem}
\usepackage{bbold}

\jyear{2021}%

\theoremstyle{thmstyleone}%
\theoremstyle{thmstyletwo}%

\theoremstyle{thmstylethree}%

\begin{document}

\title[Velocity-Dependent Forces in Quantum and Classical Fluids]{Velocity-Dependent Forces and Non-Hydrodynamic Initial Conditions in Quantum and Classical Fluids}

\author[1]{\fnm{T.~R.} \sur{Kirkpatrick}}\email{tedkirkp@umd.edu}
\equalcont{These authors contributed equally to this work.}
\author[2]{\fnm{D.} \sur{Belitz}}\email{dbelitz@uoregon.edu}
\equalcont{These authors contributed equally to this work.}


\affil[1]{\orgdiv{Institute for Physical Science and Technology}, \orgname{University of Maryland}, \orgaddress{\street{} \city{College Park}, \postcode{20742}, \state{MD}, \country{USA}}}

\affil[2]{\orgdiv{Department of Physics, Institute for Fundamental Science, and Materials Science Institute}, \orgname{University of Oregon}, \orgaddress{\street{} \city{Eugene}, \postcode{97403}, \state{OR}, \country{USA}}}




\abstract{We consider a fermionic fluid in a non-equilibrium steady state where the fluctuation-dissipation theorem is not valid
and fields conjugate to the hydrodynamic variables are explicitly required to determine response functions.
We identify velocity-dependent forces in the kinetic equation that are equivalent to such fields. They lead to driving terms in
the hydrodynamic equations and to corrections to the hydrodynamic initial conditions.}

\maketitle

\section{Introduction}
\label{sec:1}

A major complication in dealing with systems that are not in thermodynamic equilibrium is the absence of a simple fluctuation-dissipation
theorem. As a result, the linear response of the system to an external perturbation is not simply given by the appropriate correlation
function, as is the case in equilibrium. As an example, consider a fluid in a non-equilibrium steady state (NESS) characterized by a
constant temperature gradient. For a classical fluid it is known that the coupling of the temperature fluctuations, via the temperature 
gradient, to the diffusive shear velocity leads to extraordinarily long-ranged temperature correlations, with the static temperature
correlation function diverging as $1/k^4$ in the limit of vanishing wave number $k\to 0$ \cite{Kirkpatrick_Cohen_Dorfman_1982c,
Dorfman_Kirkpatrick_Sengers_1994, Ortiz_Sengers_2007}. A quantum fluid, e.g., conduction electrons in a metal, in the hydrodynamic
regime will show the same behavior since the structure of the hydrodynamic equations is the same as in a classical fluid \cite{Kirkpatrick_Belitz_2022},
while in the collisionless regime the singularity is expected to be weaker since the shear velocity is ballistic rather than diffusive.
These long-range correlations are generic in the sense that they do not require any fine tuning of the system parameters, and they
are not related to a broken symmetry. They reflect a generalized rigidity \cite{Anderson_1984} of the system that is an intrinsic
feature of the NESS \cite{Kirkpatrick_Belitz_Dorfman_2021, us_tbp}.

These correlations can be probed by light scattering, and in classical fluids this has confirmed the theoretical predictions with high
accuracy, see Ref. \cite{Sengers_Ortiz_Kirkpatrick_2016} and references therein. These are difficult experiments even in classical
fluids, due to the small scattering angles involved. In the quantum regime they would be even harder, since the fluctuations become
weaker with decreasing temperature. It therefore is desirable to probe the long-ranged correlations via response or relaxation
experiments. Since the relation between correlation functions and response functions is not clear in a NESS, this requires the
explicit consideration of external fields conjugate to the observables in question, i.e., the shear velocity and the temperature.
This can easily be done within the framework of time-dependent Ginzburg-Landau (TDGL) theory \cite{Hohenberg_Halperin_1977}.
However, the physical nature of these fields (for a response experiment) is unclear, and so is the relation to the corresponding
initial-value problem (for a relaxation experiment). 

In this paper we use kinetic theory to elucidate the nature of these fields and the related initial conditions. We show that
the fields correspond to velocity-dependent forces. They lead to an initial shear stress or momentum current 
(for the velocity-related force), or an initial heat current (for the temperature-related force), in addition to the initial shear velocity 
or temperature perturbation. These results open the door for experiments that probe the generalized rigidity of the NESS, and the 
related long-ranged correlations, via response or relaxation experiments.

\section{Kinetic Theory}
\label{sec:2}

We consider a fermionic quantum fluid; bosonic fluids or classical fluids can be treated analogously.
Let $f_{\bm p}({\bm x},t)$ be the averaged single-particle phase space distribution function, or  $\mu$-space distribution function
in the terminology of Ehrenfest \cite{Ehrenfest_1907}, which in equilibrium
is the Fermi-Dirac distribution. It is governed by the Uehling-Uhlenbeck equation \cite{Uehling_Uhlenbeck_1933,
Dorfman_vanBeijeren_Kirkpatrick_2021}
\be
\partial_t f_{\bm p}({\bm x},t) + {\bm v}_{\bm p}\cdot{\bm\partial}_{\bm x} f_{\bm p}({\bm x},t) 
     + {\bm F}_{\bm p}({\bm x},t)\cdot{\bm\partial}_{\bm p} f_{\bm p}({\bm x},t) = {\cal C}(f)_{\bm p}({\bm x},t)\ .
\label{eq:1}
\ee
Here ${\cal C}(f)$ is the collision integral that takes into account the fermionic statistics \cite{Uehling_Uhlenbeck_1933,
Landau_Lifshitz_X_1981}, and ${\bm v}_{\bm p}$ is the quasiparticle velocity. The 
electron-electron interaction is not of qualitative importance for our purposes, and we neglect it. Then the quasiparticle
energy is $\epsilon_p = {\bm p}^2/2m$, and ${\bm v}_{\bm p} = {\partial}_{\bm p}\,\epsilon_p = {\bm p}/m$,
with $m$ the free (or band, in a solid) electron mass. ${\bm F}_{\bm p}({\bm x},t)$ is a force given by
the gradient of a potential $\Phi$ that in general is velocity dependent. For our purposes it suffices to consider a
potential $\Phi$ that is a separable function of the momentum ${\bm p}$ and the space-time position $({\bm x},t)$,
\be
{\bm F}_{\bm p}({\bm x},t) = -{\bm\partial}_{\bm x}\,\Phi_{\bm p}({\bm x},t) = -\psi({\bm p})\, {\bm\partial}_{\bm x} h({\bm x},t)\ .
\label{eq:2}
\ee
The momentum-dependent function $\psi({\bm p})$ will be specified later. 

\subsection{Hydrodynamic equations}
\label{subsec:2.1}

From the Uehling-Uhlenbeck equation (\ref{eq:1}) one can derive equations for the five hydrodynamic variables, viz., 
the mass density $\rho$, the three components of the fluid velocity ${\bm u}$, and the entropy per particle $s/n$ or,
alternatively, the temperature $T$. Of the three fluid velocity components we will need only one of the two transverse
ones, $u_{\perp}$, to be specified below. The fluctuations of the five hydrodynamic variables are given in terms of momentum moments of the deviations
$\delta f_{\bm p}$ of the distribution function $f_{\bm p}$ from the equilibrium distribution $f_{\bm p}^{(0)}$,
\bse
\label{eqs:3}
\be 
\delta A_{\alpha}({\bm k},t) = \frac{1}{V}\sum_{\bm p} a_{\alpha}({\bm p})\,\delta f_{\bm p}({\bm k},t)\quad (\alpha=\rho,\perp,s)\ ,
\label{eq:3a}
\ee
where $A_{\rho}({\bm k},t) = m\,\rho({\bm k},t)$, $A_{\perp}({\bm k},t) = \rho\,u_{\perp}({\bm k},t)$, $A_s({\bm k},t) = Tn\,(s/n)({\bm k},t)$, 
and we have performed a spatial Fourier transform. The coefficients $a_{\alpha}$ are \cite{Belitz_Kirkpatrick_2022}
\be
a_{\rho}({\bm p}) = 1\quad,\quad a_{\perp}({\bm p}) = \hat{\bm k}_{\perp}\cdot{\bm p}\quad,\quad a_s({\bm p}) = \epsilon_p -\mu - sT/n\ .
\label{eq:3b}
\ee
\ese
Here $\mu$, $s$, $T$, and $n$ are the spatially averaged chemical potential, entropy density, temperature, and particle
number density, respectively, and $\rho = mn$. $\hat{\bm k}_{\perp}$ is the unit vector perpendicular to ${\bm k}$ that lies in the plane
spanned by ${\bm k}$ and the fixed temperature gradient ${\bm\nabla}T$ that characterizes the NESS, see Fig.~1, and
$u_{\perp} = \hat{\bm k}_{\perp}\cdot{\bm u}_{\perp}$, with ${\bm u}_{\perp}$ the shear velocity.
\begin{figure}[b]
\centering
\includegraphics[width=3.5cm]{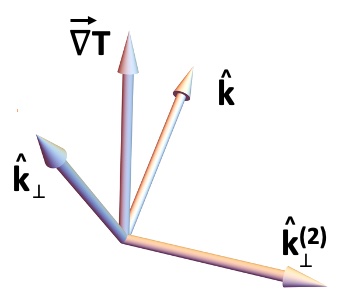}
\caption{$\hat{\bm k}_{\perp}$ is perpendicular to $\hat{\bm k}$ in the  $\hat{\bm k}\,$-$\bm{\nabla} T$ plane. $\hat{\bm k}_{\perp}^{(2)}$
              is the third of the three vectors that span the wave-vector space. }
\end{figure}

For the linearized theory one can obtain equations for the $A_{\alpha}$ by means of projection
operators \cite{Kirkpatrick_Belitz_2022}, for the nonlinear theory one can employ the Chapman-Enskog method \cite{us_tbp}.
In the absence of the force $F_{\bm p}$ the resulting Navier-Stokes equations are the same as for a classical fluid.
In an approximation that neglects, (1) pressure fluctuations, which are much faster than the diffusive fluctuations of
the temperature and the shear velocity, (2) all nonlinearities except for the crucial coupling between the
temperature gradient and the shear velocity, and further replaces all thermodynamic derivatives and transport coefficients
by their spatially averaged values, 
\footnote{This approximation neglects effects of the temperature gradient that are less leading than the one caused by
                the ${\bm\nabla}T$ term in Eq.~(\ref{eq:4b}) \cite{Kirkpatrick_Cohen_Dorfman_1982c}.}
they are
\bse
\label{eqs:4}
\bea
(\partial_t + \nu{\bm k}^2) u_{\perp}({\bm k},t) = 0\ ,
\label{eq:4a}\\
(\partial_t + \DT{\bm k}^2)T({\bm k},t) + \perpgradT u_{\perp}({\bm k},t) = 0\ .
\label{eq:4b}
\eea
\ese
Here $\nu = \eta/\rho$, with $\eta$ the shear viscosity, is the kinematic viscosity, and 
$\DT = \kappa/c_p$, with $\kappa$ the heat conductivity and $c_p$ the specific heat per volume at constant pressure, 
is the heat diffusion coefficient. We have chosen to write the heat equation in terms of the temperature fluctuations rather than
the entropy fluctuations. At constant pressure, they are proportional to each other: $\delta T({\bm x},t) = (Tn/c_p)\,\delta(s/n)({\bm x},t)$.

In what follows we will augment these equations by fields conjugate to $u_{\perp}$ and $T$, respectively, which
allows for the calculation of response functions. In order to calculate correlation functions one needs to add Langevin
forces to the equations. This has been done in Refs.~\cite{Kirkpatrick_Belitz_2022} and \cite{us_tbp}
for the equilibrium and NESS equations, respectively.

\subsection{Fields conjugate to the shear velocity and the temperature}
\label{subsec:2.2}

Now we specify the momentum dependence of the force in Eq.~(\ref{eq:2}) by writing
\be
{\bm F}_{\bm p}({\bm x},t) = \frac{-1}{\rho}\,a_{\perp}({\bm p})\, {\bm\partial}_{\bm x} h_{\perp}({\bm x},t)\ .
\label{eq:5}
\ee
Since we are only interested in the linear response of the system to the applied field $h_{\perp}$ we can neglect the
fluctuations of the distribution function in the force term in Eq.~(\ref{eq:1}). Furthermore, consistent with the simplifications
that led to Eqs.~(\ref{eqs:4}) we can replace the chemical potential by its spatial average. The distribution function in the 
last term on the left-hand side of Eq.~(\ref{eq:1}) then becomes
\bse
\label{eqs:6}
\be
f_{\bm p}^{\text{NESS}}({\bm x}) = \frac{1}{\exp{(\epsilon_p - \mu)/T({\bm x})}+1}\ .
\label{eq:6a}
\ee
Its spatial gradient at constant pressure (as opposed to constant chemical potential) is
\be
{\bm\partial}_{\bm x}\,f_{\bm p}^{\text{NESS}}({\bm x}) = w({\bm p})\,a_s({\bm p})\,\frac{1}{T}\,{\bm\nabla}T
\label{eq:6b}
\ee
with 
\be w({\bm p}) = f_{\bm p}^{(0)}[1 - f_{\bm p}^{(0)}]/T\ .
\label{eq:6c}
\ee
\ese
In the expression for $w$ we have replaced $T({\bm x})$ by
its spatial average $T$ since we already have an explicity ${\bm\nabla}T$ in Eq.~(\ref{eq:6b}).

Using these expressions in Eq.~(\ref{eq:1}) and repeating the derivation of the hydrodynamic equations we find
\bse
\label{eqs:7}
\bea
(\partial_t + \nu{\bm k}^2) u_{\perp}({\bm k},t) &=& \nu\,{\bm k}^2\,\frac{1}{\rho}\,h_{\perp}({\bm k},t)\ ,
\label{eq:7a}\\
(\partial_t + \DT{\bm k}^2)T({\bm k},t) + \perpgradT u_{\perp}({\bm k},t) &=& \perpgradT\,\frac{1}{\rho}\,h_{\perp}({\bm k},t)\ .
\label{eq:7b}
\eea
\ese
We see that the result of the field $h_{\perp}$ is a shift of the shear velocity in the viscous term
in Eq.~(\ref{eq:7a}), and the coupling term in Eq.~(\ref{eq:7b}), $u_{\perp} \to u_{\perp} - h_{\perp}/\rho$.
This is consistent with the TDGL theory known as Model H in Ref.~\cite{Hohenberg_Halperin_1977}. Note that
$h_{\perp}$ contributes to the heat equation at the level of the Euler equations (i.e., the hydrodynamic
equations for an inviscid fluid), whereas in the velocity equation it contributes to the dissipative term.
For later reference we also note that Eq.~(\ref{eq:7a}) does not contain the temperature gradient.
However, the temperature gradient leads to a coupling of the shear velocity into the heat equation via
the last term on the left-hand side of Eq.~(\ref{eq:7b}). In equilibrium such a coupling is not present.

Analogously, we can introduce a field conjugate to the temperature (or, equivalently, to the entropy per particle) by
choosing a force
\be
F_{\bm p}({\bm x},t) = \frac{-1}{c_p}\,a_s({\bm p})\,{\bm\partial}_{\bm x} h_T({\bm x},t)\ .
\label{eq:8}
\ee
This force contributes to the velocity equation at the Euler level, and to the heat equation
via a contribution to the dissipative term. We find
\bse
\label{eqs:9}
\bea
(\partial_t + \nu{\bm k}^2) u_{\perp}({\bm k},t) &=& \frac{-1}{\rho}\,\perpgradT\,h_T({\bm k},t)\ ,
\label{eq:9a}\\
(\partial_t + \DT{\bm k}^2)T({\bm k},t) + \perpgradT u_{\perp}({\bm k},t) &=& \DT {\bm k}^2\,\frac{T}{c_p}\,\,h_T({\bm k},t)\ .
\label{eq:9b}
\eea
\ese
This is again consistent with Model H in Ref.~\cite{Hohenberg_Halperin_1977}.

\section{The Physical Realization of the Forces}
\label{sec:3}

Forces of the type given by Eq.~(\ref{eq:2}) can be physically realized by means of the relation between
the linear response to an external field and the corresponding initial-value problem \cite{Forster_1975, Chaikin_Lubensky_1995}.
\footnote{It is remarkable that this relation holds in complete generality, including systems that are not in equilibrium.
                      We will elaborate on this elsewhere \cite{us_tbp}.}
Consider the force given by Eq.~(\ref{eq:5}). If such a force is adiabatically switched on, and then switched
off at time $t=0$ according to
\be
h_{\perp}({\bm k},t) = h_{\perp}({\bm k})\,e^{\epsilon t}\,\Theta(-t)\ ,
\label{eq:10}
\ee
with $\epsilon>0$ infinitesimal, it produces an initial shear velocity
\be
u^{(0)}_{\perp}({\bm k}) = \frac{1}{\rho}\,\frac{1}{V}\sum_{\bm p} (\hat{\bm k}_{\perp}\cdot{\bm p})\,\delta f_{\bm p}({\bm k},t=0) 
                                               = \frac{1}{\rho}\,h_{\perp}({\bm k})\ ,
\label{eq:11}
\ee
and an associated initial shear strain. The field $h_{\perp}$ can thus be realized by imposing an initial shear velocity on the system. 
On hydrodynamic time scales the shear strain is associated with a shear stress
\be
\sigma_{\perp}({\bm k},t) = \frac{1}{m}\,\frac{1}{V}\sum_{\bm p} ({\hat{\bm k}}\cdot{\bm p})\,(\hat{\bm k}_{\perp}\cdot{\bm p})\,
          \delta f_{\bm p}({\bm k},t) = \eta\,ik\,u_{\perp}({\bm k},t)\ .
\label{eq:12}
\ee                    
After a few mean-free times we thus have an initial shear stress
\footnote{The time scale on which Eq.~(\ref{eq:13}) is valid is an example of a `slip time',
see Ref.~\cite{Kirkpatrick_Belitz_Dorfman_2021b} and references therein.}
\be
\sigma_{\perp}^{(0)}({\bm k}) = \eta\,ik\,\frac{1}{\rho}\,h_{\perp}({\bm k})\ .
\label{eq:13}
\ee
The effective hydrodynamic initial conditions, at a microscopic time after the field $h_{\perp}$ has been switched off,
are given by Eqs.~(\ref{eq:11}) and (\ref{eq:13}). The shear velocity is in the hydrodynamic subspace spanned by 
the five hydrodynamic modes, but the shear stress is not.
To see how it enters the hydrodynamic equations we write the kinetic equation as an initial-value
problem. For simplicity we consider the linearized theory. Equation (\ref{eq:1}) yields
\be
\left(-iz + i{\bm k}\cdot{\bm v}_{\bm p} - \Lambda({\bm p})\right) \delta f_{\bm p}({\bm k},z) = \delta f_{\bm p}({\bm k},t=0)\ .
\label{eq:14}
\ee
Here $\Lambda({\bm p})$ is the linearized collision operator, and we have performed a temporal Laplace transform
with $z$ the complex frequency. We can now use a projector technique as in Ref.~\cite{Kirkpatrick_Belitz_2022}.
Let ${\cal P}$ and ${\cal P}_{\perp}$ be projectors that project onto and perpendicular to, respectively, the
hydrodynamic subspace. By operating on Eq.~(\ref{eq:14}) we can then derive hydrodynamic equations
that contain a non-hydrodynamic initial condition given by $\sigma_{\perp}^{(0)}({\bm k})$, which is a second
momentum moment of ${\cal P}_{\perp} \delta f_{\bm p}$ evaluated at an effective initial time chosen such that
Eq.~(\ref{eq:13}) is valid. The calculation yields (see Appendix~\ref{app:A} for details)
\be
(-iz + \nu{\bm k}^2)\, u_{\perp}({\bm k},z) = u^{(0)}_{\perp}({\bm k})
            - \frac{\nu}{\langle p_y v_{\bm p}^x\vert p_y v_{\bm p}^x\rangle}\,i\vert{\bm k}\vert\,\sigma_{\perp}^{(0)}({\bm k})\ , \quad
\label{eq:15}
\ee
which is equivalent to Eq.~(\ref{eq:7a}) in the same sense in which the linear-response problem in
equilibrium statistical mechanics is equivalent to an initial-value problem, see Ref.~\cite{Forster_1975}.
We will elaborate on this point in Sec.~\ref{sec:4}. Here 
$\langle A_{\bm p}\vert B_{\bm p}\rangle = (1/V)\sum_{\bm p} w({\bm p}) A_{\bm p} B_{\bm p}$
is a scalar product with $w({\bm p})$ from Eq.~(\ref{eq:6c}) as the  weight function. 

From Eq.~(\ref{eq:12}) we see
that the right-hand side of Eq.~(\ref{eq:15}) can be interpreted as an effective initial shear velocity. In the low-temperature limit,
and ignoring factors of $O(1)$, we have $\langle p_y v_{\bm p}^x\vert p_y v_{\bm p}^x\rangle \approx n\mu$
and Eq.~(\ref{eq:15}) becomes
\be
(-iz + \nu{\bm k}^2)\, u_{\perp}({\bm k},z) = \left(1+\frac{\nu k^2}{\mu}\,\frac{\eta}{n}\right)\,u^{(0)}_{\perp}({\bm k})\ .
\tag{15'}
\ee
We see that the non-hydrodynamic part of the initial condition leads to a correction that is quadratic in the
wave number, as is the viscous term on the left-hand side, but small by a factor of $\nu k^2\tau$, with $\tau$
an effective relaxation time on the order of $\eta/\mu n$. Note that Eqs.~(\ref{eq:15}) and (15') do not
contain the temperature gradient, which is consistent with Eq.~(\ref{eq:7a}), but the gradient provides a
coupling between the shear velocity and the temperature fluctuations, see the remark after Eqs.~(\ref{eqs:7}).
 
Analogously, the field $h_T$ is equivalent to imposing an initial temperature perturbation. By the same arguments
as above this leads to an effective initial heat current, which also is outside of the hydrodynamic subspace. 
This initial-value problem contains an explicit temperature gradient even in the $u_{\perp}$ equation,
consistent with Eq.~(\ref{eq:9a}).

\section{Summary, and Discussion}
\label{sec:4}

In summary, we have considered a fluid, classical or quantum, in a NESS characterized by
a constant temperature gradient. Such a NESS is known to harbor very long-ranged correlations
of the shear velocity and the temperature \cite{Kirkpatrick_Cohen_Dorfman_1982c,
Dorfman_Kirkpatrick_Sengers_1994, Ortiz_Sengers_2007}.
In classical fluids these have been observed by light scattering, see Ref.~\cite{Sengers_Ortiz_Kirkpatrick_2016}
and references therein. However, in contrast to equilibrium fluids, these correlation functions are not
related in any simple way to response functions, due to the absence of a simple fluctuation-dissipation
theorem. In order to obtain the response functions, which would be easier to measure, especially
in quantum fluids at low temperatures, one needs to explicitly consider fields conjugate to the
fluid velocity and the temperature.

We have shown that such fields are realized by velocity-dependent forces in the underlying 
kinetic equation. Deriving the hydrodynamic equations in the presence of such a force leads,
for the case of a force relevant for the shear velocity, to the driven hydrodynamic equations
(\ref{eqs:7}). In order to obtain a physical realization of such a driving field, we then considered
the kinetic equation as an initial-value problem and showed that adiabatically switching on the
field, and switching it off at an initial time $t=0$, leads to initial conditions that are equivalent
to applying the field. This allows for studying the NESS effects by means of relaxation 
experiments as an alternative to scattering experiments, and it is in analogy to the relation
between the linear response to external fields and a suitable initial-value problem in
equilibrium statistical mechanics \cite{Forster_1975}. However, the velocity-dependent
forces lead to initial conditions that are not purely hydrodynamic in nature. In the case
of the shear velocity relaxation problem, the initial condition involves an initial shear
stress in addition to an initial shear velocity, and the shear-stress contribution, which in
the space of hydrodynamic variables, involves a transport coefficient (viz., the shear
viscosity). This is expressed in Eqs.~(\ref{eq:15}) and (15'), which represent the initial-value
problem that is equivalent to the driven Eq.~(\ref{eq:7a}). An analogous discussion applies
to the field conjugate to the temperature.

We finally mention that the velocity-dependent entropic force that involves the field
conjugate to the temperature, Eq.~(\ref{eq:8}), is realized by one of the fundamental 
forces of nature, viz., the gravitational force. This was pointed out by Luttinger \cite{Luttinger_1964}
who used it to derive Kubo-type expressions for thermal transport coefficients. The underlying
physics is that, according to general relativity, heat is equivalent to a mass divided by the speed 
of light squared, and what couples to the gravitational field is the relativistic mass, which is velocity 
dependent with the leading velocity dependence being quadratic \cite{Landau_Lifshitz_II_1971, Luttinger_1964}.
As a result, the temperature of a self-gravitating body in thermo-mechanical equilibrium is higher
at the center than at the surface, see Sec.~27 in Ref.~\cite{Landau_Lifshitz_V_1980}. 
In the terrestrial gravitational field this is an exceedingly small effect. However, with parameters
as appropriate for a typical neutron star the temperature difference between the surface and the center is
on the order of $10\%$.

\backmatter

\bmhead{Acknowledgments} We thank Thomas Schaefer for a discussion.

\section*{Declarations}
\begin{itemize}
\item No datasets were generated or analyzed during the current study.
\end{itemize}

\begin{appendices}

\section{Derivation of Eq. (15)}
\label{app:A}

Here we show how to derive Eq.~(\ref{eq:15}) from Eq.~(\ref{eq:14}). We define a scalar product in the space of modes
\be
\langle A_{\bm p}\vert B_{\bm p}\rangle = (1/V)\sum_{\bm p} w({\bm p}) A_{\bm p} B_{\bm p}
\label{eq:A.1}
\ee
with $w({\bm p})$ as given after Eq.~(\ref{eq:6b}) as the  weight function. Let ${\cal L}_0$ be the hydrodynamic subspace
spanned by the hydrodynamic modes $a_{\rho}$, $a_{\perp}$, and $a_s$,
and in addition the longitudinal component of the velocity and the second transverse component, 
and define a projector ${\cal P}$ that projects onto ${\cal L}_0$ ,
\be
{\cal P} = \sum_{\alpha} \frac{\vert a_{\alpha}({\bm p})\rangle\langle a_{\alpha}({\bm p})\vert}{\langle a_{\alpha}({\bm p})\vert a_{\alpha}({\bm p})\rangle}
\label{eq:A.2}
\ee
The projector onto the complementary space is ${\cal P}_{\perp} = \mathbb{1} - {\cal P}$, with $\mathbb{1}$ the unit operator.

Now consider Eq.~(\ref{eq:14}), let ${\bm k} = (k,0,0)$, and impose initial conditions in the form of an initial shear velocity
\bse
\label{eqs:A.2.1}
\be
u_{\perp}^{(0)}({\bm k}) = \frac{1}{V} \sum_{\bm p} v_{\bm p}^y \delta f_{\bm p}({\bm k},t=0)\ ,
\label{eq:A.2.1a}
\ee
and an initial shear stress
\be
\sigma_{\perp}^{(0)}({\bm k}) = \frac{1}{V} \sum_{\bm p} p_y v_{\bm p}^x\, \delta f_{\bm p}({\bm k},t=0)\ .
\label{eq:A.2.1b}
\ee
\ese
Operating on Eq.~(\ref{eq:14}) from the left we obtain
\be
{\cal P} \delta f_{\bm p}({\bm k},t=0) = \left(-iz + {\cal P} i{\bm k}\cdot{\bm v}_{\bm p}\right) {\cal P} \delta f_{\bm p}({\bm k},z) 
          + {\cal P} i{\bm k}\cdot{\bm v}_{\bm p} {\cal P}_{\perp} \delta f_{\bm p}({\bm k},z) \ .
\label{eq:A.3}
\ee
Here we have used the projector properties $\mathbb{1} = {\cal P} + {\cal P}_{\perp}$ and ${\cal P}\Lambda({\bm p}) = 0$. 
Performing the same operation with ${\cal P}_{\perp}$ instead of ${\cal P}$ we find
\be
{\cal P}_{\perp} \delta f_{\bm p}({\bm k},t=0) = -\left[\Lambda_{\perp}({\bm p}) + O(z,k)\right] {\cal P}_{\perp}  \delta f_{\bm p}({\bm k},z)
     + {\cal P}_{\perp}  i{\bm k}\cdot{\bm v}_{\bm p} {\cal P}  \delta f_{\bm p}({\bm k},z)
\label{eq:A.4}
\ee          
where $\Lambda_{\perp}({\bm p}) = {\cal P}_{\perp} \Lambda({\bm p}) {\cal P}_{\perp}$. Note that $\Lambda_{\perp}({\bm p})$ has an
inverse since the zero eigenvalues of $\Lambda({\bm p})$ have been projected out. Solving Eq.~(\ref{eq:A.4}) for 
${\cal P}_{\perp}  \delta f_{\bm p}({\bm k},z)$, and inserting the result in Eq.~(\ref{eq:A.3}), we find
\bea
\left(-iz + {\cal P} i{\bm k}\cdot{\bm v}_{\bm p} + {\cal P}  i{\bm k}\cdot{\bm v}_{\bm p} \Lambda_{\perp}^{-1}({\bm p})
      i{\bm k}\cdot{\bm v}_{\bm p}\right)   {\cal P} \delta f_{\bm p}({\bm k},z) &=& {\cal P} \delta f_{\bm p}({\bm k},t=0) 
      \nonumber\\
      && \hskip -100pt + {\cal P} i{\bm k}\cdot{\bm v}_{\bm p} \Lambda_{\perp}^{-1}({\bm p}) \delta f_{\bm p}({\bm k},t=0)\ .
\label{eq:A.5}
\eea      
Of the two initial-condition terms on the right-hand side of Eq.~(\ref{eq:A.5}) the first one is in the hydrodynamic
subspace, but the second one is not. Furthermore, the non-hydrodynamic initial condition involves a collision
operator and hence is dissipative in nature, and Eq.~(\ref{eq:A.5}) is a closed equation for ${\cal P}  \delta f_{\bm p}({\bm k},z)$
only if one neglects the non-hydrodynamic initial condition.   

By operating on Eq.~(\ref{eq:A.5}) from the left with the hydrodynamic modes $\langle a_{\alpha}({\bm p})\vert$
we obtain the hydrodynamic equations as an initial-value problem. The only unusual aspect is the non-hydrodynamic
initial condition. The latter does not contribute to the density equation, obtained by multiplying from the left with $\langle 1\vert$,
on account of the identity 
\be
\langle i{\bm k}\cdot{\bm v}_{\bm p} \vert  \Lambda_{\perp}^{-1}({\bm p}) = 0\ ,
\label{eq:A.6}
\ee
which holds since ${\bm k}\cdot{\bm v}_{\bm p} \in {\cal L}_0$. If the initial shear stress, Eq.~(\ref{eq:A.2.1b}), is the only
non-hydrodynamic initial condition it does not contribute to the heat equation either, since the relevant matrix element
vanishes by symmetry. It does, however, contribute to the equation for the shear velocity. The relevant matrix element is
\be
M = \langle p_y\vert{\cal P}  i{\bm k}\cdot{\bm v}_{\bm p}^x \vert \Lambda_{\perp}^{-1}({\bm p})\vert \delta f_{\bm p}({\bm k},t=0)\rangle\ .
\label{eq:A.7}
\ee
If we again take the initial shear stress to be the only non-hydrodynamic initial condition we can project onto $\sigma_{\perp}$
and have
\be
M = \langle p_y v_{\bm p}^x \vert \Lambda_{\perp}^{-1}({\bm p}) \vert p_y v_{\bm p}^x \rangle\,\frac{ik}{\langle p_y v_{\bm p}^x \vert p_y v_{\bm p}^x \rangle}\,
     \langle p_y v_{\bm p}^x \vert \delta f_{\bm p}({\bm k},t=0)\rangle\ .     
\label{eq:A.8}
\ee
The first factor on the right-hand side of Eq.~(\ref{eq:A.8}) we recognize as minus the shear viscosity $\eta$, and if we remember that the 
mode $a_{\perp}({\bm p}) = p_y$ yields $\rho$ times the transverse velocity $u_{\perp}$, see Eqs.~(\ref{eqs:3}), we obtain the transverse velocity
equation in the form
\be
(-iz + \nu{\bm k}^2)\, u_{\perp}({\bm k},z) = u_{\perp}^{(0)}({\bm k}) - \frac{\nu i k}{\langle p_y v_{\bm p}^x \vert p_y v_{\bm p}^x \rangle}\,\sigma_{\perp}^{(0)}({\bm k})\ ,
\label{eq:A.9}
\ee
which is Eq.~(\ref{eq:15}). Finally, the
matrix element in the denominator is $\langle p_y v_{\bm p}^x \vert p_y v_{\bm p}^x \rangle = 8(n\mu + Ts)/3$ by
Eq.~(A.23) in Ref.~\cite{Belitz_Kirkpatrick_2022}. Ignoring the factor of $8/3$ and the $Ts$ correction to $n\mu$,
and using Eq.~(\ref{eq:12}), we obtain the transverse velocity equation in the form of Eq.~(15').

\end{appendices}




\end{document}